\documentclass[conference]{IEEEtran}
\IEEEoverridecommandlockouts
\usepackage{cite}
\usepackage{amsmath,amssymb,amsfonts}
\usepackage{algorithmic}
\usepackage{graphicx}
\usepackage{textcomp}
\usepackage{xcolor}
\def\BibTeX{{\rm B\kern-.05em{\sc i\kern-.025em b}\kern-.08em
    T\kern-.1667em\lower.7ex\hbox{E}\kern-.125emX}}
\begin{document}

%\title{Work-in-Progress: Traded Control Transfer for Managing Real-Time Sensor Uncertainties in Autonomous Vehicle}

%\title{Work-in-Progress: Traded Control Transfer for Managing Real-Time Sensor Uncertainties in Autonomous Vehicle\thanks{This paper has been peer-reviewed and accepted for presentation at the 2024 IEEE Real-Time Systems Symposium (RTSS).}}

\title{Work-in-Progress: Traded Control Transfer for Managing Real-Time Sensor Uncertainties in Autonomous Vehicle\thanks{\textbf{This paper has been peer-reviewed and accepted by the 2024 IEEE Real-Time Systems Symposium (RTSS).}}}

%\author{\IEEEauthorblockN{1\textsuperscript{st} Md Sakib Galib Sourav}
\author{\IEEEauthorblockN{Md Sakib Galib Sourav}
\IEEEauthorblockA{\textit{Department of EECS} \\
\textit{University of Toledo}\\
Toledo, USA \\
msourav@rockets.utoledo.edu}
\and
%\IEEEauthorblockN{2\textsuperscript{nd} Liang Cheng}
\IEEEauthorblockN{Liang Cheng}
\IEEEauthorblockA{\textit{Department of EECS} \\
\textit{University of Toledo}\\
Toledo, USA \\
liang.cheng@utoledo.edu}

}

\maketitle

\begin{abstract}

At Levels $2$ and $3$ of autonomous driving defined by the Society of Auto-motive Engineers, drivers must take on certain driving responsibilities, and  automated driving must sometimes yield to human control.  This situation can occur in real time due to uncertainties in sensor measurements caused by environmental factors like fog or smoke. To address this challenge, we propose a method to manage real-time sensor uncertainties in autonomous vehicles by monitoring sensor conflicts and dynamically adjusting control authority to maintain safe operation. However, to achieve this, we have introduced a novel metric called the Degree of Conflicts (DoC), which quantifies the conflict between real-time sensor data by measuring the differences between data from multiple sensors. Our approach aims to demonstrate the importance of selecting an appropriate DoC threshold for transferring control between the automation agent and the human driver.
The results have shown that choosing the correct DoC threshold can enhance safety by promptly handing over the driving control from the automation system to the human driver in challenging conditions.
\end{abstract}


\section{Introduction}

Cooperative driving, an evolution of Advanced Driver Assistance Systems (ADAS), enhances human-automation collaboration\cite{b1}. A key component is the arbitration system \cite{b2}, which allocates control between human and machine. While research has focused on automation assisting drivers \cite{b3,b4}, little attention has been given to human drivers assisting automation agents, with \cite{b5} being the first to address this issue for potentially dangerous automated actions. 

To our knowledge, no work has explored arbitration systems focusing on degraded perception in automation agents. 

Modern autonomous vehicles use various sensors, each with inherent limitations. Although sensor fusion can mitigate individual sensor weaknesses \cite{b6}, highly conflicting information may degrade performance \cite{b7}, potentially leading to disastrous consequences in autonomous vehicles. To address this critical issue, we propose a novel  traded control transfer approach for autonomous vehicles that leverages human assistance when automation perception is degraded.

%Alom 02/09: Ensure that this claim is accurate. A quick Google search could reveal similar work, which could pose a problem.

%Alom 02/09: Compromise and conflict have different meanings. "Compromise" is typically used in the context of security attacks when a system is breached. Are you using the correct term in your case? Please rewrite it accordingly.

 \begin{figure}[h]
  \centering
  \includegraphics[width=25.5em]{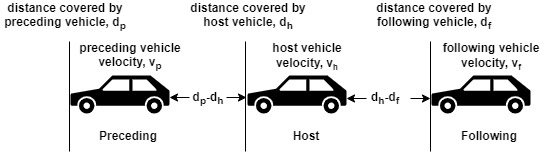}
  \caption{Car-following scenario}
\end{figure}

%Alom 02/09: I prefer using "Problem Formulation" instead of "Scenario." Could you create a separate section for Problem Formulation? Please include some mathematical equations to illustrate the mismatch in distance and explain why a solution is necessary. In the following section, present the proposed approach.

\section{Problem Statement}

We adopted a car-following scenario with a host car, a preceding car, and a following car to measure safety in terms of rear-end collision risks (Fig. 1). The host car uses radar and LiDAR, fused via Extended Kalman Filter (EKF). Selecting appropriate process noise covariance and measurement noise covariance matrices for EKF poses a significant challenge. While measurement noise covariance is typically based on sensor specifications, process noise covariance often relies on expert knowledge and trial-and-error \cite{b8}. This becomes problematic for autonomous vehicles in dynamic environments. Improper modeling of process noise covariance can exacerbate issues when sensor measurements greatly differ, potentially causing the filter to rely on inaccurate data and produce skewed state estimates. This compromises the vehicle's environmental perception and decision-making. Adaptive filtering, proposed as a solution, has limitations in real-world applications \cite{b9}, including parameter interdependence, stability issues, and sensitivity to initial conditions. %Vargas et al. \cite{b10} suggest that at least one sensor type usually maintains functionality across various situations. 
While a voting scheme might help, determining sensor accuracy remains challenging. When environmental uncertainties exceed the autonomous system's perception capabilities, control should be transferred to the human driver for safety.

%We adopted a car-following scenario with a host car, a preceding car, and a following car to measure safety in terms of rear-end collision risk (Fig. 1). The host car, equipped with radar and LiDAR, uses Extended Kalman Filter (EKF) for sensor fusion. Selecting appropriate process and measurement noise covariance matrices is challenging, with process noise often determined through trial and error [8]. Adaptive filtering, proposed as a solution, introduces its own limitations in real-world applications, including parameter interdependence, stability issues, and sensitivity to initial conditions. Balancing adaptation speed with accuracy, computational demands. These challenges can compromise the vehicle's environmental perception and decision-making capabilities. When environmental uncertainties exceed the system's capabilities, control should be transferred to the human driver for safety.

\section{Proposed Approach}

%We proposed a traded control transfer method (Arbitrator block in Fig. 2) which continuously monitors the conflicts between radar and LiDAR data and the conflicts is quantified through degree of conflicts (DoC). Arbitrator decides who will control the vehicle, automation agent or human driver based on the value of DoC.

%We proposed a traded control transfer method (Arbitrator block in Fig. 2) which continuously monitors the conflicts between radar and LiDAR data. This approach aligns with the insight from Vargas et al. \cite{b10}, who suggest that at least one sensor type usually maintains functionality across various situations. Building on this, we quantify the conflicts through a degree of conflicts (DoC) metric. The Arbitrator then decides who will control the vehicle, automation agent or human driver, based on the value of DoC. This method allows us to detect edge cases related to weather uncertainties and ensure driver safety by transferring control to the human driver when necessary. During these periods of human control, the automation system can learn safe driving behavior from the human driver, potentially improving its performance in challenging conditions. Our approach provides a dynamic way to adjust control based on relative sensor agreement, enhancing safety in autonomous driving scenarios while also creating opportunities for continuous improvement of the automated system.

We proposed a traded control transfer method (Arbitrator block in Fig. 2) that continuously monitors conflicts between radar and LiDAR data, quantified through a degree of conflicts (DoC) metric. This aligns with Vargas et al. \cite{b10}, who suggest at least one sensor type usually maintains functionality across various situations. The arbitrator decides between automation agent or human driver control based on the DoC value. This approach attempts to detect weather-related edge cases, ensures driver safety by transferring control when necessary, and allows the automation to learn from human driving behavior during these periods. Our method dynamically adjusts control based on sensor agreement, enhancing safety while enabling continuous improvement of the automated system.

%Alom 02/09: Could you please briefly explain the proposed model, outlining each section from Figure \ref{} at an abstract level?

%\section{Scenario and Proposed Approach}

%We adopted a car-following scenario involving a host car following a preceding car as shown in Fig. 1. We also included another vehicle that follows the host car. We did so to measure safety in terms of the possibility of rear-end collisions with respect to the host car and the car following it. The proposed traded control transfer approach for the host car is shown in Fig. 2. 
 
 \begin{figure}[h]
  \centering
  \includegraphics[width=22em]{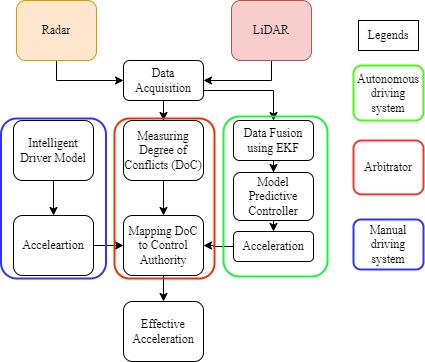}
  \caption{System architecture incorporating proposed traded control transfer method}
\end{figure}

%Alom 02/09:  I believe vehicle models should be included in the background study. We shouldn't explain background or preliminary information after the proposed model. Please move this preliminary information to follow the introduction section.

\subsection{Autonomous Driving System}

Both host car and the car following it consist of the Adaptive Cruise Control (ACC) system (different values of model parameters and MPC parameters) that maintains a desired distance and velocity with the car in front of it by adjusting the vehicle's acceleration. The simplified state-space model for both host and following vehicle can be described as follows \cite{b10}. 

%The control input \( u \) is generated based on the distance error \( \delta d \) and the velocity error \( \delta v \) \cite{b9}:

%\begin{equation}
%\begin{aligned}
%\delta d &= d - d_r \\
%\delta v &= v_p - v_h
%\end{aligned}
%\label{eq:errors}
%\end{equation}

%where \( d_r = T_\text{hw} v_h + d_0 \) represents the dynamic safe distance between the preceding and host vehicles. Here, \( T_\text{hw} \) is the time headway, \( d_0 \) is the standstill distance, and \( d \) is the relative distance between the preceding and host vehicles. Additionally, \( v_p \) and \( v_h \) denote the velocities of the preceding and host vehicles, respectively. The simplified state-space model for both host and following vehicle can be described as follows \cite{b9}:

\begin{align}
\dot{x} &= A x + B u \\
y &= C x
\end{align}

The state variables are the distance error \( \delta d \), velocity error \( \delta v \), and host vehicle acceleration \( \dot{v}_h \) ($=a_A$). The control input \( u \) is the system input, and the output \( y \) denotes the acceleration resulting from the host vehicle's traction force. The state space matrices are defined as \cite{b10}: 

\begin{tabular}{ccc}
$A = \begin{bmatrix}
0 & 1 & -T_\text{hw} \\
0 & 0 & -1 \\
0 & 0 & -\frac{1}{T_\text{e}}
\end{bmatrix}$ &
$B = \begin{bmatrix}
0 \\
0 \\
-\frac{K_\text{e}}{T_\text{e}}
\end{bmatrix}$ &
$C = \begin{bmatrix}
0 & 0 & 1
\end{bmatrix}$
\end{tabular}
Here, $T_e$ is the engine time constant and $K_e$ is the steady-state gain.
%where \( x = [\delta d, \delta v, \dot{v}_h]^T \) is the state vector, and

%\begin{align}
%A &= \begin{bmatrix}
%0 & 1 & -T_\text{hw} \\
%0 & 0 & -1 \\
%0 & 0 & -\frac{1}{T_\text{eng}}
%\end{bmatrix} \\
%B &= \begin{bmatrix}
%0 \\
%0 \\
%-\frac{K_\text{eng}}{T_\text{eng}}
%\end{bmatrix} \\
%C &= \begin{bmatrix}
%0 & 0 & 1
%\end{bmatrix}
%\end{align}

\subsection{Manual Driving System}
The Intelligent Driver Model (IDM) captures the dynamics of a vehicle based on its interactions with the preceding vehicle and the desired driving characteristics of the driver. The vehicle acceleration based on the IDM is given by \cite{b12}:
\begin{equation}
%\frac{dv_h}{dt} = a_{max} \left[1 - \left(\frac{v_h}{v_0}\right)^{\delta} - \left(\frac{s^*(v_h, \Delta v)}{s}\right)^2\right]
a_H = a_{max} \left[1 - \left(\frac{v_h}{v_0}\right)^{\delta} - \left(\frac{s^*(v_h, \Delta v)}{s}\right)^2\right]
\label{eq:idm_accel}
\end{equation}
where $v_h$ is the current velocity of the host vehicle and $\Delta v$ is the relative velocity to the preceding vehicle.
The desired minimum gap $s^*(v_h, \Delta v)$ is calculated as \cite{b11}:
\begin{equation}
s^*(v_h, \Delta v) = s_0 + \max\left(0, v_hT_{hw} + \frac{v_h\Delta v}{2\sqrt{a_{max}b}}\right)
\label{eq:idm_gap}
\end{equation}

The Intelligent Driver Model (IDM) uses several key parameters. These include $v_0$, the desired velocity in free traffic; $T_{hw}$, the desired gap to the preceding vehicle; $a_{max}$, the maximum acceleration; and $b$, the comfortable deceleration. Additionally, $\delta$ represents the acceleration exponent, which influences acceleration smoothness, while $s_0$ denotes the minimum distance between vehicles in congested traffic.

\subsection{Arbitrator} In our proposed traded control trasfer approach, the arbitration or control authority parameter $\lambda$ depends on the conflicts between LiDAR and radar.

%There are several methods \cite{b10} used to quantify the conflicts between sensors: Conflicting coefficient-based expression methods, distance-based conflict expression methods, pignistic probability-based conflict expression methods, gambling credibility distance-based conflict expression methods, Normal measure, Fuzzy measure, etc.

\subsubsection{Conflict Analysis}
We propose a measure of sensor conflicts called the Degree of Conflicts (DoC), which uses a sigmoid function to obtain a normalized value ($0$ to $1$) representing the conflict between sensors. The DoC is denoted as:
%We propose a measure of sensor conflicts called Degree of Conflicts (DoC) using a sigmoid function to obtain the normalized value (0 to 1) of the conflicting data between sensors. DoC denoted as:

%Alom02/09: I would like to use  z instead of x, as x usually refers to single values, whereas z denotes something different.

\begin{equation}
DoC = f(z)=\frac{1}{1 + e^{-10(z-1)}}
\end{equation}

%Alom 02/09: why 10 ? why not 15/20, explain.

Here, $z$ is the difference between the radar and LiDAR data averaged over a certain length of window. However, it is noteworthy that we use the sigmoid function with a scaling factor of 10 and a center of 1 for transitions in the positive domain, which fits our non-negative input z. Due to differing noise characteristics of radar and LiDAR, identical readings are unlikely. Even if they occur, at z = 0, the conflict value is (0.00005) which is negligible and indicates minimal conflict. With the increase of scaling factor, the function becomes steeper. Although it does not matter for our current method, in future our future method needs it to be less steep to ensure smooth transition of control. 

%for matching readings.

%\begin{description}
%\item[$\cdot$] $x$ is the difference between the radar and LiDAR data averaged over a certain length of window
%\item[$\cdot$] $k$ is the steepness of the curve or the tuning factor
%\item[$\cdot$] $x_0$ is the value of $x$ at the sigmoid function's center

%Our sigmoid function with steepness $10$ and center $1$ transitions in the positive domain, fitting our non-negative input $z$. Due to different noise characteristics, identical readings are unlikely. Even if that happens, at $z=0$, it yields a negligible conflict $(0.00005)$, reflecting minimal conflict for matching readings. 
%This realistically captures radar-LiDAR interactions, accounting for rare exact matches while modeling typical differences.
%\end{description}

\subsubsection{DoC-Based Control Authority Allocation}
%\subsection{Mapping DoC to Control Authority} 

In this step, we determine value for the authority parameters \( \lambda_H \) and \( \lambda_A \), which decide who controls the autonomous vehicle. It is assumed that \( \lambda_H + \lambda_A = 1 \), where \( \lambda_A \) represents the degree of authority of the automation agent. In the control transfer process, \( \lambda_A \) is defined as:

\begin{equation}
\lambda_A = \begin{cases}
1 & \text{if } 0 \leq \text{$DoC$} < \text{$DoC$}_{Th} \\
0 & \text{otherwise}
\end{cases}
\end{equation}

The vehicle operates autonomously (\( \lambda_A = 1 \)) and drives itself until the DoC reaches the threshold \( \text{DoC}_{Th} \). Beyond this threshold, the value of \( \lambda_H \) is set to 1, indicating that the human driver takes control.

%In this step, we determine discrete values of authority parameters $\lambda_H_d$, $\lambda_A_d$. These parameters decide who will be in control of the the autonomous vehicle. It is assumed that, $\lambda_H_d$ + $\lambda_A_d$ = 1 where $\lambda_A$ is the degree of authority of automation agent. In traded control transfer, $\lambda_A$ is considered to be discrete and defined as:
%\begin{equation}
%\lambda_A_d= \begin{cases}1 ; & 0 \leqslant D o C<D o C_{T h} \\ 0 ; & \text { otherwise }\end{cases}
%\end{equation}
%The car is fully autonomous ($\lambda_A_d$ = 1) and drives itself until DoC reaches a threshold ($DoC_{Th}$). Beyond that the value of $\lambda_H_d$ is set as 1 %, the human driver is notified to takeover the control and only input from human operator is considered. The threshold value of DoC can be either fixed and continuous.

 %Alom 02/07: Do you need an extra section for it ? What will you do with this value ?
 
\subsection{Effective Acceleration} The ultimate acceleration at discrete time index $k$ is calculated as follows \cite{b13}: 
\begin{equation}
    a(k)=\lambda_H a_H(k)+\lambda_A a_A(k)
\end{equation}
  
%Alom 02/07: Time index is represented as a subscript, where k denotes the value of the function, such as a(k). revise it
   
where $a$ is the effective acceleration at time step $k$, $a_H$ and $\lambda_H$ are the acceleration input from human driver and its degree of control authority, $a_A$ and $\lambda_A$ are  the acceleration input from automation system and its degree of control authority. 

%Alom 02/09: Are \( \lambda_A \) and \( \lambda_A_d \) the same parameter? If so, you mentioned earlier that \( \lambda_A_d \) is the authority parameter, whereas \( \lambda_A \) is used as the acceleration input. This is confusing. Please clarify their roles and distinctions.

% How to determine the threshold of DoC?
% What is the rationale to determine the threshold of DoC?
%  Frequency of control transfer
% consequence: number of collision, # of times lane was crossed, i.e. not in lane, or distance between vehicles surrounding the vehicle of interests

%\section{Preliminary Result}
%We tested our proposed approach in the case of car-following with and without traded-control (Fig. 3 and 4). We introduced a foggy condition between time steps of 190 and 300. This affected the LiDAR data which in turn affected the distance data fused by Kalman filter. This resulted in slowing down of the host vehicle and in turn the vehicle following it (Fig. 3). To implement traded control (Fig. 4), the value of DoC was empirically chosen to be 0.5 at which point the control is switched between the automation agent and the human driver. Once the control is transferred to human driver, they drove the car at desired speed instead of slowing it down until the control is switched to automation agent again. Then we measured compromised safety (CS) in terms of the reduction in dynamic safe distance between the host car and the car following it between the concerned time steps (N) for both with and without traded control scenarios, then calculated the improvement in safety as follows: 

\section{Experiments and results}
We tested our proposed approach for car-following scenario (Fig. 1) in MATLAB, both without and with traded control transfer (Figures 3 and 4). The experiment ran for 500 time steps (sampling time 0.1s), with the host vehicle equipped with radar and LiDAR sensors to measure the distance to the preceding vehicle. To introduce challenging conditions, we simulated fog between time steps 190-300 (between 19-30 seconds) taking the fact into consideration that fog affects the LiDAR readings but not the radar, consequently impacting the Extended Kalman Filter-fused distance data. As a result, the automation agent of the host vehicle perceived the distance to the preceding vehicle as less than the actual distance. In the scenario without traded control transfer (Fig. 3), the host vehicle responded by decelerating to slow down (between 190 and 300 time steps) to maintain a distance which is more than the safe distance from the preceding vehicle. This action reduced the distance between the host vehicle and the vehicle following it. For the proposed traded control transfer implementation (Fig. 4), we empirically set the threshold value of DoC as 0.5 for switching control between the automation agent and human driver. Under human control, the car maintained desired speed instead of slowing down until control reverted to the automation agent again.

\subsection{Evaluation Metrics} 

%In the evaluation, we consider the two performance metrics, as explained in what follows.

\begin{itemize}
    \item \textbf{Safety Improvement:} Safety is measured in terms of Compromised Safety (CS) which is the reduction in dynamic safe distance between the host car and the following car over the relevant time steps ($T$) for both scenarios.  We calculate the improvement in safety by comparing the CS values under different control strategies over these time steps. The Safety Improvement is calculated using the following equation:

%\begin{equation}
%\begin{aligned}
%& \text{Safety Improvement (SI)} = \\
%& \frac{1}{T} \sum_{k=1}^T \frac{\text{CS}^{\text{No Traded control}}_t - %\text{CS}^{\text{Traded control}}_t}{\text{CS}^{\text{No Traded control}}_t} \right\\
%&
%\end{aligned}
%\end{equation}

\begin{equation}
\begin{aligned}
& \text{Safety Improvement (SI)} = \\
& \frac{1}{T} \sum_{k=1}^T \frac{\text{CS}^{\text{No Traded control}}_t - \text{CS}^{\text{Traded control}}_t}{\text{CS}^{\text{No Traded control}}_t}
\end{aligned}
\end{equation}

SI measures the average reduction in compromised safety when transitioning from the "No Traded control" scenario to the "Traded control" scenario over the \( T \) time steps. %A positive value indicates an improvement in safety, while a negative value suggests a decrease in safety performance.

    \item \textbf{Redundant Human Engagement (RHE):} RHE is defined as below: \begin{equation}
\text{RHE} = \frac{\text{Total time when IDM is active}}{\text{Total time when fog is absent}}
\end{equation}We
calculated the RHE for different thresholds as well. 
%Alom 02/09: This defination is not clear. revise it, add the mathmetical equation  of this matric. 

\end{itemize}

%Alom 02/09: To ensure consistency, always use t to represent time steps. Avoid using different symbols such as k or N. This will help maintain clarity and uniformity in your notation throughout the document.

\subsection{Result Discussion}

In this subsection, we analyze the experimental results we perform to evaluate the  effectiveness of the proposed approach. 

%Alom 02/09:  Added new section results  discussion: Here explain both Table results here. 

\begin{table}[htbp]
\caption{Safety Improvement and Redundant Human Engagement at Different Thresholds}
\begin{center}
\begin{tabular}{|c|c|c|}
\hline
\textbf{Threshold $({DoC}_{Th})$} & \textbf{SI (\%)} & \textbf{RHE} (\%) \\
\hline
0.2 & 100 & 100 \\
\hline
0.5 & 80.79 & 0 \\
\hline
0.8 & 10.23 & 0 \\
\hline
\end{tabular}
\label{tab:combined_metrics}
\end{center}
\end{table}

%   \begin{table}[htbp]
%\caption{Safety Improvement at different thresholds}
%\begin{center}
%\begin{tabular}{|c|c|}
%\hline
%\textbf{Threshold $({DoC}_{Th})$} & \textbf{Safety Improvement (\%)} \\
%\hline
%0.2 & 100 \\
%\hline
%0.5 & 80.79 \\
%\hline
%0.8 & 10.23 \\
%\hline
%\end{tabular}
%\label{tab:human_engagement}
%\end{center}
%\end{table}

%Alom 02/09: explain the table results here. The table illustrates the relationship between thresholds and Safety Improvement (%), write here

%\begin{table}[htbp]
%\caption{Redundant Human engagement at different thresholds}
%\begin{center}
%\begin{tabular}{|c|c|}
%\hline
%\textbf{Threshold $({DoC}_{Th})$} & \textbf{Redundant Human Engagement} \\
%\hline
%0.2 & 100\% \\
%\hline
%0.5 & 0\% \\
%\hline
%0.8 & 0\% \\
%\hline
%\end{tabular}
%\label{tab:human_engagement}
%\end{center}
%\end{table}    

%These preliminary results indicate that our traded-control approach improves safety while varying human engagement based on the chosen threshold. The 0.5 threshold effectively balanced automation and human control in foggy conditions.

 \begin{figure}[h]
  \centering
  \includegraphics[width=20em]{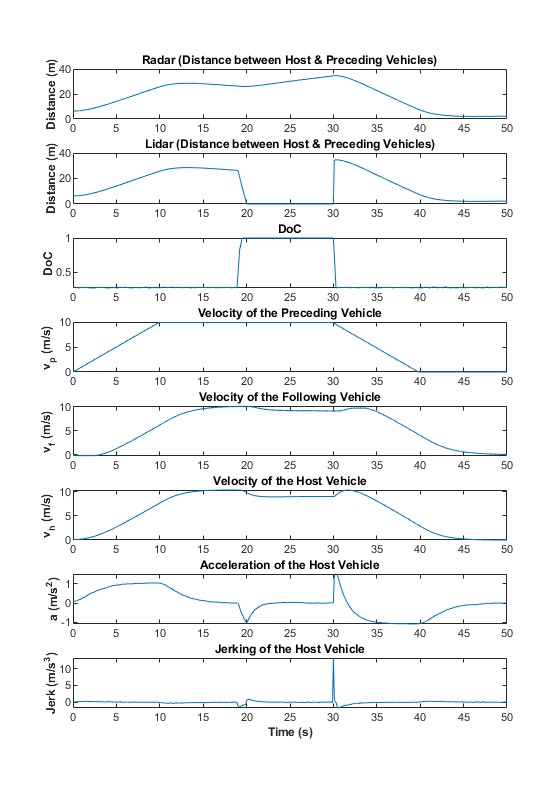}
  \caption{Car-following without traded control transfer.}
\end{figure}

 \begin{figure}[h]
  \centering
  \includegraphics[width=20em]{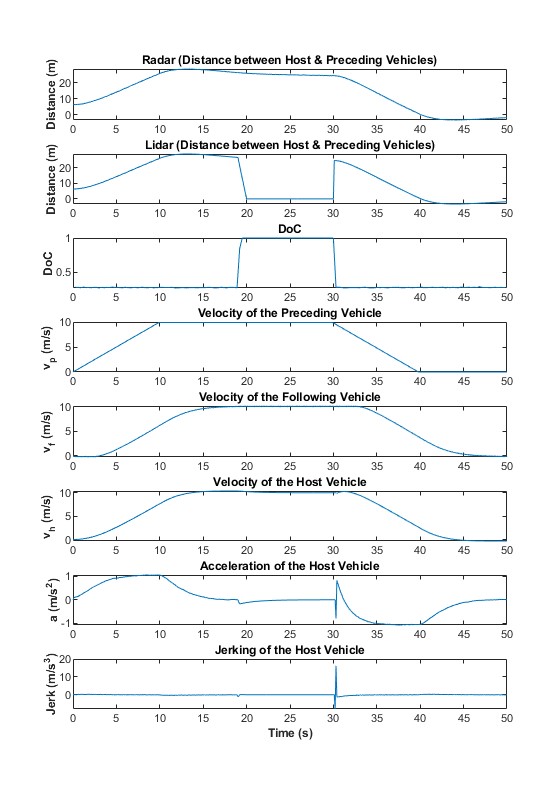}
  \caption{Car-following with traded control transfer.}
\end{figure}

The data Table \ref{tab:combined_metrics} demonstrates an inverse relationship between safety improvement and redundant human engagement across thresholds. While 0.2 maximizes safety at 100\% but requires full human involvement, 0.8 eliminates unnecessary engagement but reduces safety improvement to 10.23\%. The 0.5 threshold emerges as the optimal balance, achieving 80.79\% safety improvement without redundant human engagement, effectively handling adverse weather condition optimizing both safety and operational efficiency. However, improvements in comfort and transition smoothness are needed. 
%Our preliminary results show the proposed approach enhances safety while optimizing human engagement. Table \ref{tab:combined_metrics} illustrates how threshold choice impacts involvement: at 0.2, 100\% engagement indicates unnecessary intervention, while at 0.5, 0\% engagement suggests effective handling of foggy conditions without unwarranted human involvement. However, improvements in comfort and transition smoothness are needed. The current traded control approach faces limitations in abrupt transitions, causing jerking (Fig. 4), and relies on manual threshold setting.

\section{Future Work}

While our proposed traded control transfer approach shows promise, several areas require further research:

\begin{itemize}
 
 \item \textbf{Vehicle Model:} Incorporate detailed vehicle model instead of simplified one used in this work.

    \item \textbf{Smooth Control Transfer:} Develop methods for smoother transitions between automation and human control to eliminate jerking ensuring comfort (Fig. 4).

    \item \textbf{Adaptive Threshold:} Design an arbitrator that utilizes multiple information sources (e.g., driver readiness, how fast the following vehicle is approaching, how fast the host vehicle is slowing down, etc.) in addition to DoC for real-time decision-making, eliminating reliance on predefined $\text{DoC}_\text{Th}$.

    \item \textbf{Human Driver Modeling:} Incorporate visibility factors into the IDM-based human driver model, adjusting outputs to reflect reduced visibility in foggy conditions for more realistic post-transfer behavior analysis.

    \item \textbf{Emergency Fallback:} Implement a system to handle scenarios where the human driver is not ready to take control when the arbitrator initiates a transfer, ensuring continuous safe operation.

    \item \textbf{Heterogeneous Following Scenarios:} Our current evaluation considers the following vehicle to be controlled by an automation agent as host vehicle. Analyzing the performance of the proposed method in scenarios, where the following vehicle is driven by a human with varying characteristics (e.g., aggressiveness, cautiousness), would provide insights into its robustness in real-world conditions.

    \item \textbf{Comparative Study:} Compare the performance of our DoC metric with other sensor data conflict measurement methods, such as those discussed in \cite{b14}.
\end{itemize}

\section{Conclusion}

This study proposes a traded control transfer method for managing sensor uncertainties in autonomous vehicles operating in adverse conditions. By continuously monitoring sensor disagreement and dynamically adjusting control authority, our approach demonstrates enhanced safety compared to fully autonomous systems. Future research will focus on smooth control transfer, adaptive threshold, visibility of human driver, emergency fallback and heterogeneous following cenarios.
%integrating shared control strategies, exploring complex scenarios, assessing driver readiness, addressing current limitations, and validating the method's effectiveness. 

%With further refinements, the proposed approach has the potential to contribute to the development of robust and reliable autonomous driving systems for safe deployment in real-world conditions. 

%This work proposes a traded control transfer method for managing sensor uncertainties in autonomous vehicles operating in adverse conditions. By continuously monitoring sensor disagreement and dynamically adjusting control authority, our approach demonstrates enhanced safety compared to fully autonomous systems. Future research will explore shared control, complex scenarios, and driver readiness as well as driver's comfort to address current limitations and validate the method's effectiveness. With further refinements, this approach can contribute to the development of robust and reliable autonomous driving systems for safe real-world deployment. 

\vspace{12pt}
\color{red}


\begin{thebibliography}{00}
\bibitem{b1} M. Marcano, S. Daz, J. Prez, and E. Irigoyen, “A Review of Shared Control for Automated Vehicles: Theory and Applications,” IEEE Transactions on Human-Machine Systems, vol. 50, no. 6, pp. 475–491.

\bibitem{b2} D. P. Losey, C. G. McDonald, E. Battaglia, and M. K. O’Malley, “A Review of Intent Detection, Arbitration, and Communication Aspects of Shared Control for Physical Human–Robot Interaction,” Applied Mechanics Reviews, vol. 70, no. 010804, Feb. 2018.
.
\bibitem{b3} J. Sarabia, M. Marcano, J. Pérez, A. Zubizarreta, and S. Diaz, “A review of shared control in automated vehicles: System evaluation,” in Frontiers in Control Engineering, Feb. 2023, p. 1058923.
.
\bibitem{b4} X. Wu, C. Su, and L. Yan, “Human–Machine Shared Steering Control for Vehicle Lane Changing Using Adaptive Game Strategy,” Machines, vol. 11, no. 8, Art. no. 8, Aug. 2023.
.
\bibitem{b5} Y. Wu, H. Wei, X. Chen, J. Xu, and S. Rahul, “Adaptive Authority Allocation of Human-Automation Shared Control for Autonomous Vehicle,” Int.J Automot. Technol., vol. 21, no. 3, pp. 541–553, Jun. 2020.

\bibitem{b6} Y. Zhang, A. Carballo, H. Yang, and K. Takeda, “Perception and sensing for autonomous vehicles under adverse weather conditions: A survey,” ISPRS Journal of Photogrammetry and Remote Sensing, vol. 196, pp. 146–177, Feb. 2023, doi: 10.1016/j.isprsjprs.2022.12.021.

\bibitem{b7} C. Thomas and N. Balakrishnan, “Mathematical analysis of sensor fusion for intrusion detection systems,” in 2009 First International Communication Systems and Networks and Workshops, Bangalore, India: IEEE, Jan. 2009, pp. 1–10.


\bibitem{b8} S. Akhlaghi, N. Zhou and Z. Huang, "Adaptive adjustment of noise covariance in Kalman filter for dynamic state estimation," 2017 IEEE Power \& Energy Society General Meeting, Chicago, IL, USA, 2017, pp. 1-5.

\bibitem{b9} A. Almagbile, J. Wang, and W. Ding, “Evaluating the Performances of Adaptive Kalman Filter Methods in GPS/INS Integration,” Journal of Global Positioning Systems, 2017.



\bibitem{b10} J. Vargas, S. Alsweiss, O. Toker, R. Razdan, and J. Santos, “An Overview of Autonomous Vehicles Sensors and Their Vulnerability to Weather Conditions,” Sensors (Basel), vol. 21, no. 16, p. 5397, Aug. 2021.


%\bibitem{b10} N. K. Ure, M. U. Yavas, A. Alizadeh, and C. Kurtulus, “Enhancing Situational Awareness and Performance of Adaptive Cruise Control through Model Predictive Control and Deep Reinforcement Learning,” in 2019 IEEE Intelligent Vehicles Symposium (IV), Jun. 2019, pp. 626–631. 

\bibitem{b11} T. Takahama and D. Akasaka, “Model Predictive Control Approach to Design Practical Adaptive Cruise Control for Traffic Jam,” IJAE, vol. 9, no. 3, pp. 99–104, 2018. 


\bibitem{b12} M. Treiber, A. Hennecke, and D. Helbing, “Congested Traffic States in Empirical Observations and Microscopic Simulations,” Phys. Rev. E, vol. 62, no. 2, pp. 1805–1824, Aug. 2000.

\bibitem{b13} M. Omae, T. Fujioka, N. Hashimoto, and H. Shimizu, “The Application of RTK-GPS and Steer-by-Wire Technology to the Automatic Driving of Vehicles and An Evaluation of Driver Behavior,” IATSS Research, vol. 30, no. 2, pp. 29–38, Jan. 2006.

\bibitem{b14} Z. Zhang, T. Liu, D. Chen, and W. Zhang, “Novel Algorithm for Identifying and Fusing Conflicting Data in Wireless Sensor Networks,” Sensors (Basel), vol. 14, no. 6, pp. 9562–9581, May 2014.




.


\end{thebibliography}
\end{document}